\documentclass[aip,
  twoside,
  twocolumn,
  floatfix,
  reprint,
  citeautoscript,
]{revtex4-1}

\usepackage{graphicx}
\usepackage{amsmath}
\usepackage{subfigure}
\usepackage{color} 
\usepackage[]{units}

\newcommand{\Sr}[0]{\text{Sr}}

\newcommand{\La}[0]{\text{La}}
\newcommand{\Br}[0]{\text{Br}}
\newcommand{\Ce}[0]{\text{Ce}}

\newcommand{\eV}[0]{\text{eV}}
\newcommand{\meV}[0]{\text{meV}}
\newcommand{\AAA}[0]{\text{\AA}} 

\newcommand{\fig}[1]{Fig.~\ref{#1}}

\newcommand{\tab}[1]{Table~\ref{#1}}

\renewcommand{\epsilon}[0]{\varepsilon}
\newcommand{\mydot}[0]{\bullet}

\newcommand{\stokes}[1]{}

\newcommand{\myscale}{0.65}

\begin{document}

\preprint{1$^{\text{st}}$ draft}

\title{
  Origin of resolution enhancement by co-doping of scintillators:\\
  Insight from electronic structure calculations 
}

\author{Daniel {\AA}berg}
\email{aberg2@llnl.gov}
\affiliation{Physical and Life Sciences Directorate, Lawrence Livermore National Laboratory, Livermore, California 94550, USA}

\author{Babak Sadigh}
\affiliation{Physical and Life Sciences Directorate, Lawrence Livermore National Laboratory, Livermore, California 94550, USA}

\author{Andr{\'e} Schleife}
\affiliation{
Department of Materials Science and Engineering, University of Illinois at Urbana-Champaign, Urbana, Illinois 61801, USA}

\author{Paul Erhart}
\email{erhart@chalmers.se}
\affiliation{Chalmers University of Technology, Department of Applied Physics, Gothenburg, Sweden}

\begin{abstract}
It was recently shown that the energy resolution of Ce-doped LaBr$_3$ scintillator radiation detectors can be crucially improved by co-doping with Sr, Ca, or Ba. Here we outline a mechanism for this enhancement on the basis of electronic structure calculations. We show that ({\em i}) Br vacancies are the primary electron traps during the initial stage of thermalization of hot carriers, prior to hole capture by Ce dopants; ({\em ii}) isolated Br vacancies are associated with deep levels; ({\em iii}) Sr doping increases the Br vacancy concentration by several orders of magnitude; ({\em iv}) $\Sr_\La$ binds to $V_\Br$ resulting in a stable neutral complex; and ({\em v}) association with Sr causes the deep vacancy level to move toward the conduction band edge. The latter is essential for reducing the effective carrier density available for Auger quenching during thermalization of hot carriers. Subsequent de-trapping of electrons from $\Sr_\La-V_\Br$ complexes then can activate Ce dopants that have previously captured a hole leading to luminescence.
This mechanism implies an overall reduction of Auger quenching of free carriers, which is expected to improve the linearity of the photon light yield with respect to the energy of incident electron or photon.
\end{abstract}

\maketitle

Scintillator radiation detectors have many applications in nuclear and radiological surveillance, high-energy physics, and medical imaging \cite{Rod97, Kno10}. The energy resolved detection of radiation is of particular interest as it enables for example the identification of fissile materials \cite{NelGosKna11}. According to counting statistics the resolution increases with luminosity, which usually results from a higher conversion efficiency, {\it i.e.}, relatively more photons are generated per incident energy. In practice the resolution is further limited by the non-linear response of the scintillator to the energy of the incident radiation \cite{Dor10}. It is usually accepted that the resulting non-proportionality arises from the competition between non-radiative quenching, defect carrier trapping, as well as activator capture and subsequent emission \cite{Dor05, Vas08, KerRosCan09, BizMosSin09, PayMosShe11}.

One of the most promising materials for detector performance is Ce-doped LaBr$_3$\cite{LoeDorEij01}. It yields an energy resolution of 2.7\%\ at a photon energy of 662\,keV in combination with an extremely fast scintillation pulse. LaBr$_3$ has been very well characterized both experimentally \cite{DorLoeVin06, *DotMcGHar07} and theoretically \cite{BizDor07, *Sin10, CanChaBou11, AbeSadErh12}. The prospect of improving energy resolution by co-doping LaBr$_3$:Ce with Sr or Ba was first noted experimentally by Yang {\it et al.} \cite{YanMenBuz12} Later Alekhin {\it et al.} revisited this aspect and using Ca and Sr achieved an improvement of energy resolution down to 2.0\%\ at 662\,keV \cite{AleHaaKho13}. A more comprehensive investigation including both the alkaline as well as earth-alkaline series revealed that better performance is only achievable when using the heavier elements of the latter series (Sr, Ca, Ba) \cite{DorAleKho13, AleBinKra13}. Several possible mechanisms were tentatively proposed to rationalize these observations: \cite{AleWebKra14}
({\it i}) reduction of the nonradiative recombination rate,
({\it ii}) an increase of the so-called escape rate of the carriers from the quenching region, or
({\it iii}) an increase in the trapping rate of Ce$^{3+}$.
Here we address this question via first principles calculations of thermodynamic and electronic properties of intrinsic and extrinsic defects as well as their complexes in Ce and Sr-doped LaBr$_3$. It is found that Br vacancies are present in LaBr$_3$ regardless of the chemical boundary conditions and are associated with a deep electronic level below the conduction band minimum (CBM). They bind to Sr, which is preferentially incorporated substitutionally on La sites. Upon complexation, the vacancy defect level moves closer to the CBM becoming more shallow. As trapping on lattice defects effectively reduces the free carrier density available for Auger quenching \cite{GriUceBur13}, we argue that this trap is essential for the improved linearity of LaBr$_3$:Ce. Our analysis is supported by the identification of distinct optical signatures associated with Ce sites in close correspondence to recent measurements \cite{AleWebKra14}. The presence of less deep trap levels also provide a rationale for the experimentally observed longer life-times \cite{AleWebKra14}.

\begin{figure}
  \centering
  \includegraphics[scale=\myscale]{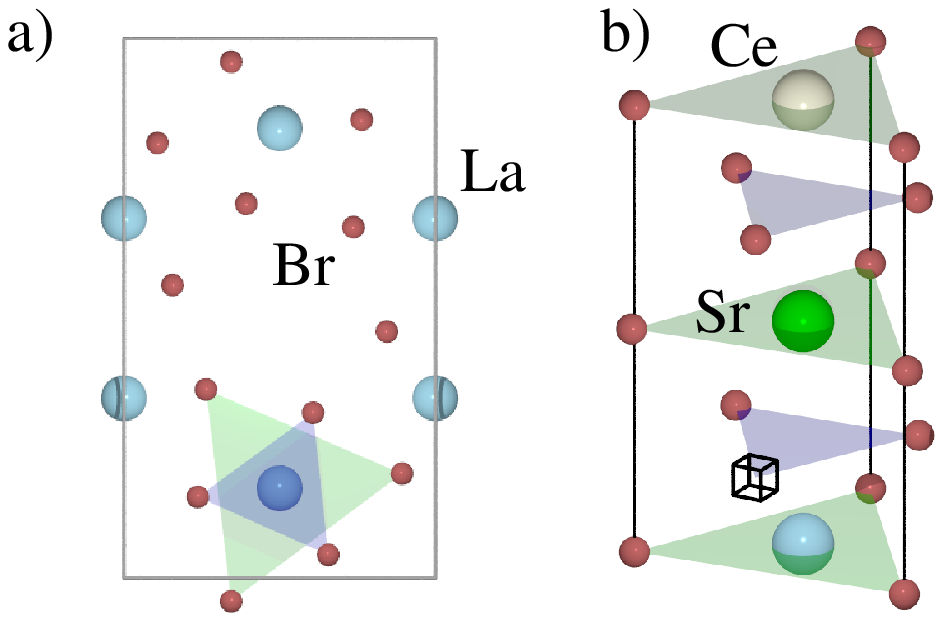}
  \caption{
   (a) View along $z$-axis of the conventional LaBr$_3$ cell. (b) Strontium-vacancy defect cluster $\Sr_\La-V_\Br$. The vacancy is illustrated by the hollow black box. For reference, we here also display one possible cerium site.
  }
  \label{fig:struct}
\end{figure}

Calculations were performed within density functional theory (DFT) using the projector augmented wave (PAW) method \cite{Blo94, *KreJou99} as implemented in the Vienna ab-initio simulation package.\cite{KreHaf93, *KreHaf94, *KreFur96a, *KreFur96b} Exchange-correlation were treated within the generalized gradient approximation.\cite{PerBurErn96} DFT+$U$ type on-site potentials \cite{DudBotSav98} were included for both La-$4f$ ($U_{\text{eff}}=10.3\,\eV$) and Ce-$4f$ states ($U_{\text{eff}}=1.2\,\eV$) in order to obtain the correct ordering of La-$5d$ and $4f$ states and to reproduce experimental Ce-$4f$ ionization energies \cite{CzySaw94, CanChaBou11, AbeSadErh12}. The plane-wave energy cutoff was set to 230\,eV and Gaussian smearing with a width of 0.1\,eV was used to determine the occupation numbers. Excited Ce-$4f^0$ states were obtained in a similar, albeit more flexible and automatized, approach as used by Canning and co-workers \cite{CanChaBou11}. First, a subspace of Ce-$4f$ states was determined by projection of single particle wave functions on spherical harmonics within the PAW spheres. This is possible for localized atomic states, such as the rare-earth 4$f$ states. The 4$f$ occupation number can then be controlled by introducing a separate electron chemical potential for the subspace. 
Lanthanum bromide adopts a hexagonal lattice structure in space group 176 (P$6_3$/m) with La and Br ions occupying Wyckoff sites $2c$ and $6h$, respectively. The calculated lattice parameters are $a=8.140\,\AAA$ and $c=4.565\,\AAA$ to be compared with experimental values of $a=7.9648(5)\,\AAA$ and $c=4.5119(5)\,\AAA$, respectively, see \fig{fig:struct} \cite{KraSchSch89}.
Defects were modeled using 168-atom supercells. $\Gamma$-point sampling was found to be sufficient to converge defect formation energies to better than 0.05\,eV. Configurations were relaxed until ionic forces were less than 10\,\meV/\AA. Defect formation energies were calculated using the formalism described in Refs.~\onlinecite{ZhaNor91, *ErhAbeLor10}. Potential alignment as well as periodic image charge corrections were taken into account to correct for finite size effects \cite{MakPay95, LanZun08}. Defect concentrations were obtained using the calculated formation energies on the basis of a self-consistent solution of the charge neutrality condition \onlinecite{ErhAlb08, ErhAbeLor10}.

In agreement with earlier calculations \cite{CanChaBou11} we find that in its ground state Ce preferentially substitutes for La with small distortions and adopts a neutral charge state corresponding to a Ce$^{3+}$-$4f^15d^0$ configuration. By choice of the DFT+$U$ parameters, the occupied $4f$ level is located 0.9\,eV above the valence band maximum (VBM). The excited Ce-$4f^05d^1$ state, which is obtained by enforcing the deoccupation of Ce-$4f$ levels, is associated with the emergence of electronic levels very close to the CBM. They are predominantly of Ce-$5d$ character and strongly hybridized with the neighboring La-$5d$ states.

\begin{figure}
  \centering
  \includegraphics[scale=\myscale]{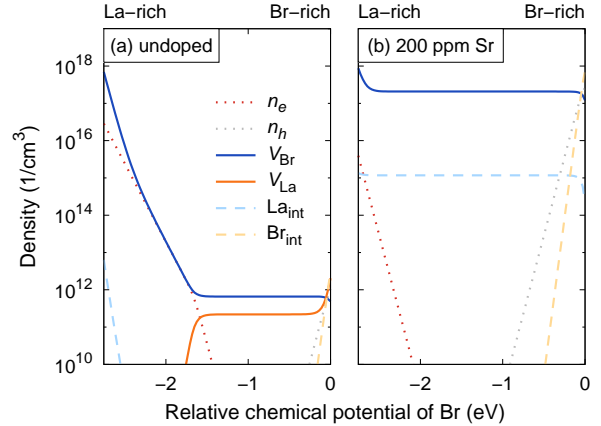}
  \caption{
    Equilibrium defect and charge carrier concentrations as a function of the Br chemical potential at a temperature of 600\,K.
    Results are shown both for (a) pure material and (b) LaBr$_3$ doped with 200 ppm Sr corresponding to the experimental doping conditions. Note that doping with Sr increases the Br vacancy concentration by several orders of magnitude.
    (Ce was not explicitly included in these figures since it does not affect the charge neutrality condition).
  }
  \label{fig:defconc}
\end{figure}

\begin{figure}
  \centering
  \includegraphics[scale=\myscale]{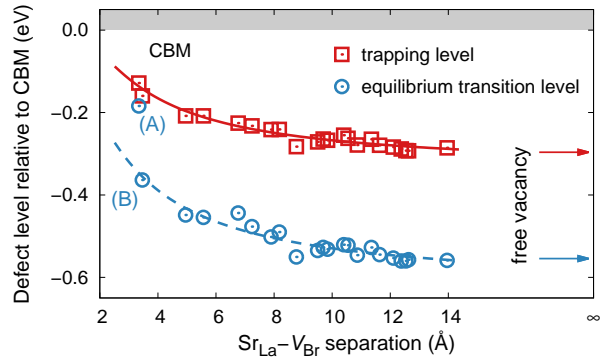}
  \caption{
    Trapping and equilibrium transition levels for the (+1/0) transition of $\Sr_\La-V_\Br$ as a function of $\Sr_\La-V_\Br$ separation.
    The former was calculated by considering the transition level for fixed ionic positions starting from the $V_\Br^\mydot$ configuration.
    The equilibrium transition level, on the other hand, was computed allowing full relaxation in both charge states.
    The data points labelled A and B indicate out-of-plane (A) and in-plane (B) configurations of nearest neighbor defect complexes, compare \fig{fig:struct}.
  }
  \label{fig:energylevels}
\end{figure}

From an extensive exploration of intrinsic defects one obtains bromine vacancies $V_\Br$ to be the most energetically favorable donor-type defect under both La and Br-rich conditions, see \fig{fig:defconc}(a). The Br vacancy is associated with an equilibrium transition level ($+1/0$) $0.55\,\eV$ below the conduction band minimum (CBM), see \fig{fig:energylevels} and a trap state 0.3\,eV below the CBM.
The associated electronic level is located inside the band gap and has La-$5d$ character. It can act as an efficient electron trap, effectively removing carriers from the light-generation process during the instrumentation pulse shape-time.

\begin{figure}
  \centering
  \includegraphics[scale=\myscale]{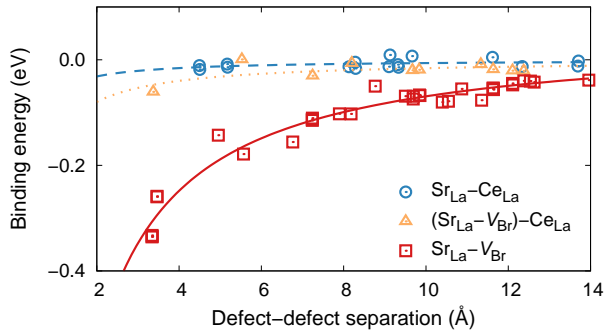}
  \caption{
    Binding energy as a function of defect separation for various defect associates.
    While there is a strong attraction between $V_\Br^\mydot$ and $\Sr_\La'$, the interaction of $\Ce_\La^\times$ with other defects is weak.
    The variation in the binding energies is related to the anisotropy of the crystal structure.
  }
  \label{fig:binding}
\end{figure}

The inclusion of strontium is accomplished by substitution on lanthanum sites. The resulting $\Sr_\La'$ defect acts as a shallow acceptor with a vanishingly small lattice distortion. Assuming a Sr concentration of 200\,ppm the concentration of Br vacancies will increase by several orders of magnitude compared to pristine or Ce-only doped material as shown in \fig{fig:defconc}.
The opposite charge states of $V_\Br^\mydot$ and $\Sr_\La'$ cause a mutual attraction, which is quantified in \fig{fig:binding} revealing a binding energy of $-0.3\,\eV$ for the nearest neighbor $(\Sr_\La-V_\Br)^\times$ complex \footnote{We here adopt the convention that negative binding energies indicate attraction.}. See \fig{fig:struct}b for an example of complex geometry.
A closer inspection of the electronic structure of the complex reveals that both the trap and equilibrium transition levels, which are associated with the Br vacancy, shift closer to the CBM by approximately 0.2\,eV compared to the isolated vacancy, see \fig{fig:energylevels}.

The shift of the defect level can be rationalized by considering that $\Sr_\La'$ introduces a point charge-like electrostatic potential that shifts the {\em local} energy scale. Localized states such as the $V_\Br$ defect level are sensitive to this shift whereas the delocalized states that make up the valence and conduction bands are unaffected, causing an effective upward shift of the vacancy level.

We can now consider the effect of the $\Sr_\La-V_\Br$ complex on the electronic structure of $\Ce_\La$. Even though the binding between $\Ce_\La$ and $\Sr_\La-V_\Br$ is weak as shown in \fig{fig:binding}, the large concentration of Ce (5\%) used experimentally \cite{YanMenBuz12, AledeHKho13} implies that the average separation between $\Ce_\La$ and $\Sr_\La-V_\Br$ is only about 2.5 unit cells. As a result, each $\Sr_\La-V_\Br$ complex will have a Ce atom in its vicinity. One can thus expect the spectroscopic signatures of cerium in Sr-doped LaBr$_3$:Ce to be affected. To this end we have calculated Ce ($4f\leftrightarrow 5d$) excitation and emission energies for isolated $\Ce_\La$ as well for various complexes of $\Ce_\La$ with $\Sr_\La-V_\Br$, where the latter are nearest neighbors. Spin-orbit interaction was not included self-consistently but rather added as a perturbation to the $4f$ states according to 
\begin{align*}
 \Delta E_{\text{so}} = 
 \begin{cases}
   -2\xi_{4f} & j =  \nicefrac{5}{2} \\
   \nicefrac{3}{2}\xi_{4f} & j = \nicefrac{7}{2},
 \end{cases}
\end{align*}
where $\xi_{4f}=0.1$\,eV as obtained from the $4f$ splitting in a Ce-$4f^05d^1$ configuration.

The thus obtained optical signatures can be categorized as follows (also compare \tab{tab:optics}):
Isolated Ce is associated with the largest excitation energy and a substantial Stokes shift. For configurations, in which the Br vacancy is a {\em first neighbor} of Ce a pronounced reduction of both excitation and emission energies is observed with a typical value of $-0.34$\,eV given in \tab{tab:optics}. Furthermore the Stokes shift is reduced compared to isolated Ce. There are several configurations with similar optical signatures. Given the accuracy that can be expected from the present DFT calculations, however, we abstain from a more detailed differentiation of these complexes and in \tab{tab:optics} only include the results for a representative cluster.
Finally, configurations, for which Ce and $V_\Br$ are {\em not} first nearest neighbors behave similar to isolated Ce.
\begin{table}
  \centering
  \caption{
    Comparison of calculated and experimental data for Ce excitation and emission ($4f^15d^0 \leftrightarrow 4f^05d^1$).
    Values in brackets in the excitation column indicate the shift with respect to free $\Ce_\La$, which is identical to site I in the case of the experimental data.
    Two values are given in the emission column corresponding to final states of $^2F_{5/2}$ and $^2F_{7/2}$, respectively. Experimental data from Ref.~\onlinecite{AleWebKra14}.
    Note that the band gap error of DFT manifests itself in a systematic underestimation of all excitation and emission energies.
  }
  \label{tab:optics}
  \begin{ruledtabular} \begin{tabular}{llccc}
      \multicolumn{2}{l}{Site} & Excitation &  Emission & Stokes shift \\
      & & (eV) & (eV) & (eV)\tabularnewline
      \hline\\[-9pt]
      \multicolumn{4}{l}{Calculation} \tabularnewline
      & $\Ce_\La$                 & 3.56 &  3.13 / 2.78 & 0.43 \tabularnewline
      & $(\Sr_\La-V_\Br)-\Ce_\La$  & 3.22 ($-0.34$)  &  2.78 / 2.43 & 0.31 \tabularnewline[6pt]
      \multicolumn{4}{l}{Experiment (Ref.~\onlinecite{AleWebKra14})} \tabularnewline
      & I             & 4.03           & 3.47 / 3.19 & 0.56 \tabularnewline
      & II            & 3.59 ($-0.44$) & 3.36 / 3.10 & 0.24 \tabularnewline
      & III           & 3.47 ($-0.56$) & 3.27 / 3.10 & 0.21 \tabularnewline
%      Site & Excitation &  Emission & Stokes shift \\
%      & (eV) & (eV) & (eV)\tabularnewline
%      \hline 
%      $\Ce_\La    $ &  3.56 &  3.13 (2.78) & 0.43 \tabularnewline
%      $(\Sr_\La-V_\Br)-\Ce_\La$  & 3.22  &  2.78 (2.43) & 0.31 \tabularnewline
%      I             & 4.03 & 3.47 (3.19) & 0.56 \tabularnewline
%      II            & 3.59 & 3.36 (3.10) & 0.24 \tabularnewline
%      III           & 3.47 & 3.27 (3.10) & 0.21 \tabularnewline
  \end{tabular} \end{ruledtabular} 
\end{table}

We now wish to point out several important facts. Firstly, we note that the shift in the calculated excitation energy ($0.34\,\eV$) between $\Ce_\La$ and $(\Sr_\La-V_\Br)-\Ce_\La$ is close to the experimental shift between sites I and II/III ($0.44/0.56\,\eV$). Thus we associate site I with $\Ce_\La$ and II/III with nearest neighbor triple complexes. Since Ce is a nearest neighbor of a Br vacancy in these configurations, the deep trap level associated with the vacancy will to a large part consist of Ce-$5d$ states. This explains, the reduction of the excitation energy for this complex compared to isolated $\Ce_\La$. Furthermore, if we assume the $4f$ level to be unaffected by the neutral complex $\Sr_\La-V_\Br$, it implies that we can identiy the shift in excitation energy with the trap depth.

To summmarize, we demonstrated that co-doping of LaBr$_3$:Ce with strontium gives rise to a shallow acceptor substituting on a lanthanum site. Overall charge neutrality requires formation of one oppositely charged bromine vacancy for each strontium atom, resulting in a steep increase of Br concentration compared to the undoped material. Moreover, the two defects are electrostatically attracted to each other and will form stable complexes that act as electron traps. We note that neutral Ce atoms are not likely to capture electrons, since the $5d$ states in this case are situated inside the conduction band \cite{CanChaBou11}. The occupied Ce $4f$ level is located rather deep inside the gap (0.6--0.9\,eV) and will therefore have a low but finite hole capture rate. If we assume that Ce is activated by sequential capture of a hole and electron this implies that hole capture is the rate limiting step. Since Auger recombination, which has been shown to be the major quenching channel at this time scale for halide scintillators \cite{GriUceBur13}, has a cubic dependence on the electron and hole densities the carrier population will be greatly reduced by non-linear quenching causing an overall non-proportional response. If, however, $\Sr_\La-V_\Br$ traps are active during the initial thermalization stage (2--10\,ps in halide systems) they will effectively reduce the free electron density. As a result, a larger density of holes will remain available for ionization of cerium activators without being quenched by the Auger mechanism. Although calculation of electron capture cross-sections for the complexes are beyond the scope of this letter, we note that a very fast capture is indeed possible. For example, in picosecond optical absorption experiments it is shown that energy transfer to europium activators in SrI$_2$:Eu may be as fast as 400 fs.\cite{UceBizBur14} 

Another time-scale of importance is the de-trapping rate from the Br-vacancy sites. As alluded to earlier, each defect complex will be in close proximity to a Ce atom. Once any of the nearby Ce atoms captures a hole, Coulombic attraction serves as a driving force for transferring the elecron from the complex to the activator. This suggests that non-linear quenching is reduced at the cost of longer decay-times; in fact, two of the three cerium sites discussed by Alehkin are associated with very long decay times ranging from 60-2500\,ns while accounting for 20-45\,\%\ of the total light output.\cite{AleWebKra14} 

We acknowledge fruitful discussions with S. Payne, G. Bizarri, and R. T. Williams.  This work was performed under the auspices of the U.S. Department of Energy by Lawrence Livermore National Laboratory under Contract DE-AC52-07NA27344 with
support from the National Nuclear Security Administration Office of Nonproliferation Research and Development (NA-22). 
P.E. acknowledges funding from the {\em Area of Advance -- Materials Science} at Chalmers. Com\-puter time allocations by the Swedish National Infrastructure for Computing at NSC (Link\"oping) and C3SE (Gothenburg) are gratefully acknowledged.
 
%\bibliography{lit1,lit2} 

\end{document}